\documentclass[pra,amsfonts]{revtex4} 

\usepackage[]{graphicx}  
\usepackage{bm}
\usepackage{amssymb}
\usepackage{amsmath}    
\usepackage{latexsym}   
\usepackage{times}
\usepackage{color}
\evensidemargin=\oddsidemargin
\def\defeq{\buildrel\rm def\over =}
\def\boxit#1{\vbox{\hrule\hbox{\vrule\kern12pt \vbox{\kern12pt#1\kern12pt}\kern12pt\vrule}\hrule}} 
\def\mcl{\mskip-6mu}  
\def\half{\textstyle{\frac{1}{2}}}
\newcommand{\ket}[1]{\mbox{$| #1 \rangle$}}  
\newcommand{\bra}[1]{\mbox{$\langle #1 |$}}  

\begin{document}

\title{On statements of experimental results expressed in the\\ mathematical
  language of quantum theory}
\author{John M. Myers}
\affiliation{Harvard University, School of Engineering and Applied Sciences,\\ 
60 Oxford Street, Cambridge, Massachusetts 02138, USA}

\author{F. Hadi Madjid}
\affiliation{82 Powers Road, Concord, Massachusetts 01742, USA}

\begin{abstract}

We note the separation of a quantum description of an experiment into a
statement of results (as probabilities) and an explanation of these results
(in terms of linear operators).  The inverse problem of choosing an
explanation to fit given results is analyzed, leading to the conclusion
that any quantum description comes as an element of a family of related
descriptions, entailing multiple statements of results and multiple
explanations. Facing this multiplicity opens avenues for exploration and
consequences that are only beginning to be explored.  Among the
consequences are these: (1) statements of results impose topologies on 
control parameters, without resort to any quantum explanation;
(2) an endless source of distinct explanations forces an open cycle of
exploration and description bringing more and more control parameters into
play, and (3) ambiguity of description is essential to the concept of
invariance in physics.
\end{abstract}

\maketitle

\flushbottom
\section{Introduction}\label{sect:1}

When a telescope projects stars of the night sky onto points of a
photograph, stars at large and small distances pile up on a single
point of the photograph.  Indeed such a ``pile-up,'' which
makes the distance to stars ambiguous, is a mathematical property of any
mapping of a space of larger dimension to a space of lesser
dimension.  Here we report on a ``piling-up'' that occurs when quantum
theory serves as mathematical language in which to describe experiments.

How does one employ quantum theory to describe experiments with
devices---lasers and lenses, detectors, \emph{etc}.\ on a laboratory
bench? One assumes that the devices generate, transform, and measure
particles and/or fields, expressed one way or another as linear operators,
such as density operators and detection operators.  In case of a
finite-dimensional quantum description, these operators are matrices.
Here we omit discussing how one arrives at the particles, in order to
focus directly on the operators that end up expressing the devices. These
operators are functions of the parameters by which one describes control
over the devices.  It is by making explicit the experimental
parameters---which we picture as \emph{knobs}---that the ambiguity of a
pile-up will become evident.

It is important to recognize that quantum theoretic descriptions of
experiments come in two parts: (1) statements of results of an experiment,
expressed by probabilities of detections as functions of knob settings,
and (2) explanations of how one thinks these results come about, expressed
by linear operators, also as functions of knob settings.  The two parts
are connected by a mapping, namely the \emph{trace}.  As one learns in
courses on quantum mechanics, given an explanation as a density operator
and a positive operator valued measure (POVM), taking the trace of the
product of the operators gives the probabilities that constitute a
statement of results.  Of special interest here is the ``inverse problem''
that stems from the assumption in quantum mechanics that experimental
evidence for quantum states is, at best, limited to probabilities of
detections.  The inverse problem amounts to finding the inverse of the
mapping defined by the trace: given a statement of results, the problem is
to determine all the explanations that generate it.  It is here that the
pile-up of the trace as a mapping impacts quantum physics.

Note that while our discussion gives knobs a prominent expression absent
in text books on quantum mechanics, we employ the standard quantum
mechanics of Dirac and von Neumann \cite{dirac,vN}, augmented only by
positive-operator-valued measures, now in widespread use.

\section{Formulation}\label{sect:2}

\begin{figure}[t]
\centerline{\includegraphics[width=3.625in]{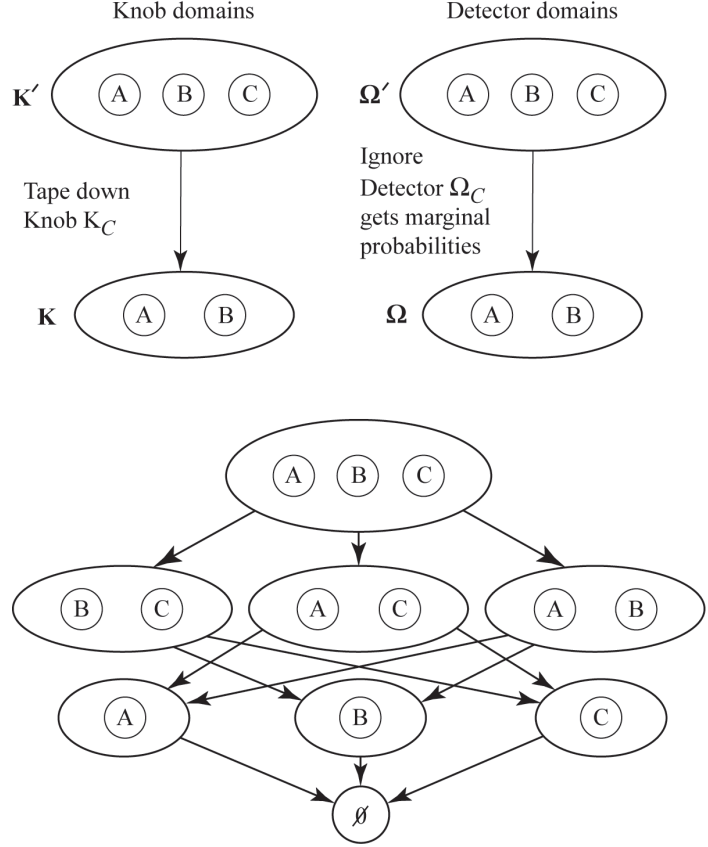}}
\caption{Lattices of domains.}\label{fig:1}
\end{figure}

We speak of the parameters by which a description expresses control over
an experiment as \emph{knobs}, with the image in mind of the physical
knobs by which an experimenter moves a translation stage or rotates a
polarization filter.  We think figuratively of hand motions by which we
configure an experiment also as knob settings.  In the mathematical
language in which we describe experimental trials, actual or anticipated,
we express any one knob by a set of \emph{settings} of the knob.  We start
with the simplest case in which each knob has a finite number of
settings. Let $K_A$, $K_B$, \emph{etc}.\ denote knobs, each of which can
be set in any of several positions.  $\#K_A$ denotes the number of knob
settings in $K_A$, \emph{etc}. When several knobs are involved, we call
all of them together a \emph{knob domain}.  For example if knobs $K_A$ and
$K_B$ are involved, then we have a knob domain $\bm{K}$ and an element of
$\bm{k}\in\bm{K}$ has the form $\bm{k} = (k_A,k_B)$ with $k_A \in K_A$ and
$k_B \in K_B$. For the number of possible settings we then have the
product: $\#\bm{K}= \#K_A\,\#K_B$.  If knob domain $\bm{K}'$ includes all
the knobs that contribute to knob domain $\bm{K}$, then we write
$\bm{K}\le\bm{K}'$; in other words knob domains form a distributive
lattice under inclusion \cite{tyler}, illustrated in Fig.~\ref{fig:1}.

Similarly we consider detectors that display one of a finite number of
outcomes. Such a detector $\Omega_A$ is a set, and $j_A\in \Omega_A$ is a
particular outcome. As with knobs, we deal with sets of detectors which we
call \emph{detector domains}, written boldface \emph{e.g.}\ as
$\bm{\Omega}$.  Detector domains also form a distributive lattice \cite{tyler}.

Experiments come in families and so do descriptions, and so do the knob
domains and detector domains that enter descriptions.  The lattice of knob
domains and the lattice of detector domains underpin expressing relations
in these families.  For example, a description involving a knob domain
$\bm{K}'$ might be simplified by fixing one knob---``taping it down'' so
to speak.  Or a description involving a detector domain $\bm{\Omega}'$
might be simplified by ignoring detector $\Omega_C$, leading to marginal
probabilities.

\begin{figure}[t]
\centerline{\includegraphics[width=3in]{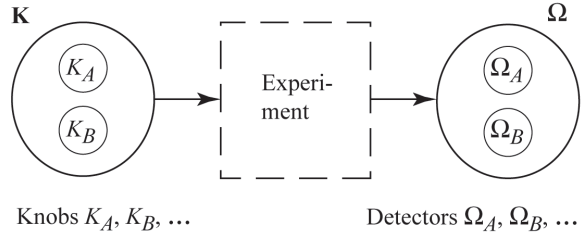}}
\caption{Quantum theory states experimental results as a parametrized 
probability measure $\mu:\ \mathbf{K}\to \mathrm{PrMeas}(\bm{\Omega})$.}\label{fig:2}
\end{figure}

A statement of (experimental) results, as expressed in quantum theory
consists of the probability of outcome $\bm{j}\in \bm{\Omega}$ for each
setting $\bm{k}$ of the knobs of $\bm{K}$, as illustrated in Fig.~\ref{fig:2}.  
We write $\mu(\bm{k},j)$ for this probability, and the probability function
$\mu:\bm{K}\times \bm{\Omega}\rightarrow [0,1]$ is what we call a
\emph{parametrized probability measure}, that is, we have that for each
$k_A\in K_A$, $k_B\in K_B$,\ \ $\mu(k_A,k_B,-):\bm{\Omega}\rightarrow[0,1]$
is a probability measure on the set of detector outcomes.  The
quantum-mechanical form of experimental reports is that of a parametrized
probability measure.

For a given knob domain $\bm{K}$ and detector domain $\bm{\Omega}$, let
PPM$(\bm{K},\bm{\Omega})$ be the space of all parametrized probability
measures.  When the number of knob settings and detections is finite, so
is the dimension of this space.  As illustrated in Sect.~\ref{sect:3} for toy
descriptions with finite numbers of knob settings and possible outcomes,
parametrized probability measures constitute points of a function space
that will play the part of a photograph onto which a larger space is
mapped.  Any $\mu \in \mbox{PPM}(\bm{K},\Omega)$ corresponds to a point on
a photographic plate.

\subsection{Explanations} 

A statement of results $\mu\!:\!\bm{K}\times\bm{\Omega}\rightarrow [0,1]$
says nothing about how its probabilities come about; that is the job of
the explanatory part of a quantum description.  An explanation of a
statement of results consists of linear operators on some Hilbert space
$\mathcal{H}$ as functions of the knob settings, including detection
operators involving $\bm{\Omega}$.  Products, tensor products, sums,
exponentiations, etc.\ of operators are combined to form a triple
$(\mathcal{H},\rho,M)$ in which $\rho$ and $M$ are functions on $\bm{K}$.
The function $\rho\!:\!\bm{K}\rightarrow \{\mbox{density operators on
}\mathcal{H}\}$ can be called a parametrized density operator, and the
function $M\!:\!\bm{K}\times\bm{\Omega}\rightarrow \{\mbox{Detection
operators on } \mathcal{H}\}$ is a parametrized positive operator-valued
measure (POVM); more precisely, for each $\bm{k} \in \bm{K}$, it is the
case that $M(\bm{k},-)\!:\!\bm{\Omega}\rightarrow \{\mbox{Detection
operators on } \mathcal{H}\}$ is a POVM on $\bm{\Omega}$.  The situation
for continuous outcome spaces calls for the technicalities of measurable
subsets, noted elsewhere \cite{tyler}.

The explanations over a given knob domain $\bm{K}$ and detector domain
$\bm{\Omega}$ consist in the set 
\begin{equation}
  \mbox{Expl}(\bm{K},\bm{\Omega})\defeq\{(\mathcal{H},\rho,M)\}
\end{equation}
ranging over
all $\mathcal{H}$, $\rho$ and $M$ of the form just stated.  
It is this space of explanations that turns out to be
larger than the space of statements of results.  

So defined, any explanation implies a statement of results \emph{via} the
familiar trace rule
\begin{equation}
  (\forall \bm{k} \in \bm{K}, \bm{\omega} \in \bm{\Omega})\quad
  \mu(\bm{k},\bm{\omega}) =\mbox{Tr}_{\mathcal{H}}[\rho(\bm{k})M(\bm{k},\bm{\omega})],
\end{equation}
where $\bm{\omega} \in \bm{\Omega}$ is an outcome. Often we
abbreviate this by
\begin{equation}
  \mu = \mbox{Tr}_{\mathcal{H}}(\rho M).
\end{equation}

As we shall see, explanations, like ``the stars in the sky,'' are space of
high dimension.  The counting of degrees of freedom for explanations is a
little involved, because we want to distinguish explanations that
have conflicting natural extensions to larger knob domains.  For this we
introduce a notion of metric deviation, to which we now turn.

\subsection{Metric deviation}

Some differences among quantum explanations ``make no difference.'' For
example if one explanation can be transformed into the other by the same
unitary transformation applied both to the density operator and the POVM,
the two explanations can be called ``unitarily equivalent.''  We are
interested in descriptions that imply the same probabilities but that have
more or less ``natural'' extensions to larger domains of knobs that, over
the extended domain, conflict in their implied probabilities.  For this it
turns out to be handy to have a notion of \emph{metric deviation}, which
to our knowledge is a novelty.  Before defining it, we start by recalling
some operator metrics.

For density operators on a common Hilbert space $\mathcal{H}$, we use the
metric defined by
\begin{equation}
\mbox{distance}[\rho(\bm{k}_1),\rho(\bm{k}_2)] =
\half\mbox{Tr}_\mathcal{H}|\rho(\bm{k}_1)-\rho(\bm{k}_2)|,
\end{equation}
where for any bounded operator $A$, $|A| \defeq \sqrt{A^\dag A}$.
We choose this metric for density operators because it determines the
least probability of error for deciding between two density operators on
the basis of probabilities of outcomes \cite{helstrom}.   

While the trace metric and other operator metrics work only for operators
on a common Hilbert space, another measure allows comparison of two
parametrized density operators defined on different Hilbert spaces.  Given
$\rho\!:\!\bm{K}\rightarrow\mbox{DensOp}(\mathcal{H})$ and
$\rho'\!:\!\bm{K}\rightarrow\mbox{DensOp}(\mathcal{H}')$, allowing that
$\mathcal{H}'$ can differ (even in dimension) from $\mathcal{H}$, we
define the \emph{metric deviation} of $\rho$ and $\rho'$ by
\begin{equation}
  \mbox{MetDev}(\rho,\rho')\defeq\sup_{\bm{k}_1,\bm{k}_2\in\bm{K}}
\half\bm{|}\mbox{Tr}_\mathcal{H}|\rho(\bm{k}_1)-\rho(\bm{k}_2)|-
\mbox{Tr}_{\mathcal{H}'}|\rho'(\bm{k}_1)-\rho'(\bm{k}_2)|\bm{|}. 
\end{equation}
In case $\mbox{MetDev}(\rho,\rho')= 0$, we call $\rho$ and $\rho'$
\emph{metrically equivalent}; otherwise $\rho$ and $\rho'$ are 
\emph{metrically inequivalent}.

\noindent\textbf{Remark}: More familiar than metric equivalence is the
notion of unitary equivalence, \emph{e.g.} in the sense that $\rho$ and
$\rho'$ are unitarily equivalent if and only if there exists a unitary
operator $U$ independent of $\bm{k}$ such that $(\forall \bm{k}\in
\bm{K})\ \rho'(\bm{k})= U\rho(\bm{k})U^\dag$.  Unitary equivalence implies metric
equivalence, but not the converse: two parametrized density operators can
be metrically equivalent without being unitarily equivalent.  Example:
$\rho(\bm{k}) = \mbox{diag}(1/2-a_k,1/2+a_k),\ 0 <a_k< 1$; then there exists
$\delta>0$ such that $\rho'(\bm{k})$ is same form with $a'_k =a_k+\delta$.  The
$\delta$ cancels out of $\rho(\bm{k})-\rho(\bm{k}')=\rho'(\bm{k})-\rho'(\bm{k}')$, so the
trace distance is invariant under the addition of $\delta$, demonstrating
metric equivalence without unitary equivalence.  (This addition of
$\delta$ to $a_k$ is not a unitary transformation because it changes the
eigenvalues.)  Addition of an increment works also, and independently, for
off-diagonal elements of generic density operators, where by
\emph{generic} we mean to rule out the special case of a unit eigenvalue
or a zero eigenvalue.

Now turn to POVMs.  Lifting the usual norm for bounded operators on a
Hilbert space $\mathcal{H}$ to parametrized bounded operators gives
a distance measure for parametrized POVMs over $\bm{K}\times\bm{\Omega}$
\begin{equation}
  \mbox{distance}[M(\bm{k}_1,-),M(\bm{k}_2,-)] \defeq \sup_{\bm{\omega} \subset
  \bm{\Omega}}\|M(\bm{k}_1,\bm{\omega})-M(\bm{k}_2,\bm{\omega})\|_\mathcal{H}.
\end{equation}
For two POVMs $M$ and $M'$ (which can differ in both their Hilbert spaces
$\mathcal{H}$ and $\mathcal{H}'$, respectively and in their outcome
domains $\bm{\Omega}$ and  $\bm{\Omega}'$) we define
\begin{equation}
\mbox{MetDev}(M,M')\defeq \sup_{\bm{k}_1,\bm{k}_2\in\bm{K}}\left|\sup_{\bm{\omega} \subset
  \bm{\Omega}}
\|M(\bm{k}_1,\bm{\omega})-M(\bm{k}_2,\bm{\omega})\|_\mathcal{H}-\sup_{\bm{\omega}' \subset
 \bm{\Omega}'}\|M'(\bm{k}_1,\bm{\omega}')-M'(\bm{k}_2,\bm{\omega}')\|_{\mathcal{H}'}\right|.
\end{equation}

\section{Dimension counting for explanations and results}\label{sect:3} 

To pursue the analogy of ``stars to photographs'' we consider toy
descriptions for which some interesting dimensions are finite.  Let
$\#\bm{K}$ be the number of settings of a knob domain $\bm{K}$, and let
$\#\bm{\Omega}$ be the number of elementary outcomes for a detector domain
$\bm{\Omega}$. (For a detector domain $\bm{\Omega}$ constituted of $n$
binary detectors, we have $\#\bm{\Omega}= 2^n$.)  Let $n_\mathcal{H}$ be
the complex dimension of a finite-dimensional Hilbert space $\mathcal{H}$,
so that a vector in $\mathcal{H}$ has $n_\mathcal{H}$ complex-valued
components.  Then the explanations $\{(\mathcal{H},\rho,M)\}$ involving
the Hilbert space $\mathcal{H}$ form (at least locally) a real manifold of
\begin{eqnarray}
  \dim[\mbox{Expl}(\bm{K},\bm{\Omega})|_\mathcal{H}]&\mcl =\mcl &
  \#\bm{K}[n_\mathcal{H}^2(\#\Omega-1)+n^2_\mathcal{H}-1]= \#\bm{K}(n_\mathcal{H}^2\,\#\bm{\Omega}-1).
\end{eqnarray}
In contrast, the dimension of statements of results (``photographic
plate'') is
\begin{equation}
  \dim[\mbox{PPM}(\bm{K},\bm{\Omega})]= \#\bm{K}(\#\bm{\Omega}-1).
\end{equation}
Subtracting the latter from the former, we find the space
of explanations on $\mathcal{H}$ of a given statement of results has
\begin{eqnarray}
  \dim[\mbox{Tr}_{\mathcal{H}}^{-1}(\mu)]&\mcl = \mcl&
  \dim[\mbox{Expl}(\bm{K},\bm{\Omega})|_\mathcal{H}]-
  \dim[\mbox{PPM}(\bm{K},\bm{\Omega})]\nonumber \\ &\mcl = \mcl&
  \#\bm{K}[n_\mathcal{H}^2(\#\bm{\Omega}-1)+n^2_\mathcal{H}-1]
  -\#\bm{K}(\#\bm{\Omega}-1)\nonumber
  \\ &\mcl = \mcl& \#\bm{K}\,\#\bm{\Omega}(n_\mathcal{H}^2-1).
\end{eqnarray}

Thus there are lots of explanations of any given parametrized probability
measure.  The next point is that among these are metrically inequivalent
explanations, as follows.  For this paragraph, by \emph{class} we mean an
equivalence class on $\mbox{Tr}_\mathcal{H}^{-1}(\mu)$ defined by
\begin{equation}
  (\mathcal{H},\rho,M)\equiv (\mathcal{H},\rho',M')\;\Leftrightarrow\;
\mbox{MetDev}(\rho,\rho')=\mbox{MetDev}(M,M')=0.
\end{equation}
By \emph{quotient space} we mean the quotient space of
$\mbox{Tr}_\mathcal{H}^{-1}(\mu)$ by this equivalence relation.  The
dimension of this quotient space is equal to the number of independent
constraints imposed by the metric equivalence.  For each pair of values
$(\bm{k}_1,\bm{k}_2)$, the demand for metric equivalence places one
constraint on parametrized density operators.  For each pair of values of
$(\bm{k}_1,\bm{k}_2)$ and each of the $\#\bm{\Omega}-1$ independent
elements of the detector domain, the demand for metric equivalence places
one constraint on detection operators.  Altogether the number of
constraints, independent or not, is given by
\begin{equation}
  \#(\mbox{constraints}) = \#\bm{K}(\#\bm{K}-1)\#\Omega/2,
\end{equation}
from which we conclude that
\begin{eqnarray}
  \dim(\mbox{quotient space})&\mcl =\mcl&
\min(\#(\mbox{constraints}),\dim[\mbox{Tr}_\mathcal{H}^{-1}(\mu)])  
  \nonumber \\ &\mcl=\mcl&
\min[\#\bm{K}(\#\bm{K}-1)\#\Omega/2,\#\bm{K}\,\#\bm{\Omega}
(n_\mathcal{H}^2-1)].
\end{eqnarray}
Distinct points of this quotient space correspond to mutually metrically
inequivalent explanations of a given parametrized probability measure, and
we have just shown that there is an infinite set of mutually metrically
inequivalent explanations. While we have shown this explicitly only for
toy cases of finite numbers of knob settings and outcomes, the set of 
metrically inequivalent explanations of a given parametrized probability
measure only gets larger when continuous sets of knob settings and outcomes
are considered \cite{aop05}.

\section{Avenues to explore}\label{sect:4}

As Sam Lomonaco remarked, the showing of ``multiple explanations'' is
analogous to the elementary proposition that through any countable set of
points runs an infinitude of curves.  From that point of view, what
we have found is in a sense no surprise.  Yet to accept that we live
in a world of multiple, inequivalent explanations is to enter a new
world, ready for exploration.  Below are four examples.

\subsection{Results without explanations imply topologies on knob domains}
Given that results can narrow possible explanations only to an infinite
set, we wondered what results alone, without any choice of explanation,
implied for the physics of knobs.  One thing that results in the form of a
parametrized probability measure imply is a topology on the knob domain.
That is, starting with a knob domain $\bm{K}$ as a set without assuming
a topology on it, any statement of results
$\mu\!:\!\bm{K}\rightarrow\mbox{PrMeas}(\bm{\Omega})$ implies a topology
$\tau_\mu$ on $\bm{K}$ that makes no assumption of any explanation:
\begin{equation}
 \tau_\mu = \{U\subset \bm{K}|(\exists V\mbox{ open in
 PrMeas}(\bm{\Omega}))\quad U= \mu^{-1}(V)\}, 
\end{equation}
(where $\mu^{-1}(V) = \{\bm{k}\in\bm{K}|\mu(\bm{k},-)\in V\}$).
If $\mu$ is an injection into $\mbox{PrMeas}(\bm{\Omega})$, then
the (bounded) uniform metric on $\mbox{PrMeas}(\bm{\Omega})$
induces a bounded metric on $\bm{K}$ \cite{tyler}.  If it is not injective, then $\mu$
induces a bounded metric on the quotient set of equivalence classes
$\bm{K}/E_\mu$ where $E_\mu$ is the equivalence relation defined by
\begin{equation}
  \bm{k}_1E_\mu \bm{k}_2 \Leftrightarrow \mu(\bm{k}_1,-)=\mu(\bm{k}_2,-).
\end{equation}
This metric, which we denote by $d_\mu$, is defined by
\begin{equation}
  d_\mu([\bm{k}_1],[\bm{k}_2])\defeq\sup_{\bm{\omega}
  \subset\bm{\Omega}}|\mu(\bm{k}_1,-)-\mu(\bm{k}_2,-)|.
\end{equation}

For $\mu \in \mbox{PPM}(\bm{K},\bm{\Omega})$ and $\mu' \in
\mbox{PPM}(\bm{K},\bm{\Omega}')$ we add to our catalog of metric
deviations by defining
\begin{equation}
  \mbox{MetDev}(\mu,\mu')\defeq \sup_{\bm{k}_1,\bm{k}_2\in \bm{K}}\bm{\left|}
|\mu(\bm{k}_1,-)-\mu(\bm{k}_2,-)|-|\mu'(\bm{k}_1,-)-\mu'(\bm{k}_2,-)|\bm{\right|}.
\end{equation}
If their metric deviation is zero,
then $\mu$ and $\mu'$ induce the same topological and
metric structures on $\bm{K}$ \cite{tyler}:
\begin{equation}
  \mbox{MetDev}(\mu,\mu') = 0 \Rightarrow E_\mu=E_{\mu'},\; 
\tau_\mu = \tau_{\mu'}, \mbox{ and }\, d_\mu = d_{\mu'}.
\end{equation}

Examples of equivalence classes of knobs relevant to entangled states that
violate Bell inequalities are discussed elsewhere \cite{ppm07}.  When $\mu$
is not injective, the coarse topology $\tau_\mu$ on $\bm{K}$ induced by
$\mu$ can be replaced by a finer topology by recognizing a finer level of
description that augments the detector domain by adding another detector, 
as discussed below in connection with equivalence classes that
characterize invariance.

\subsection{Endless cycle of extensions of inequivalent descriptions}\label{subsect:4.2}

\begin{figure}[t]
\centerline{\includegraphics[width=4.875in]{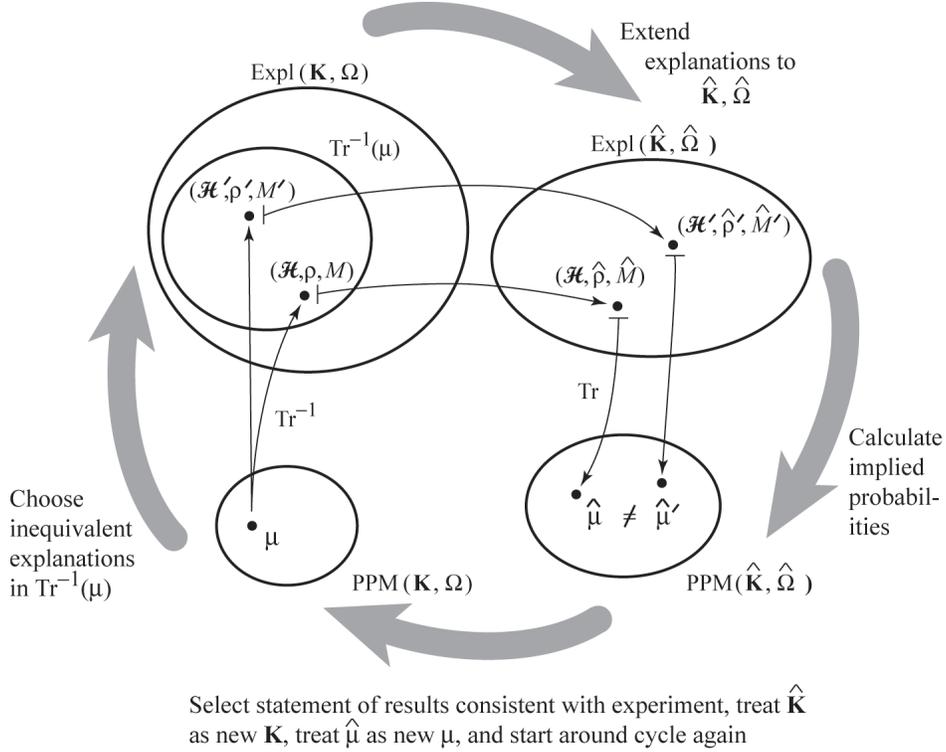}}  
\caption{Expanding cycle of results and explanations.}\label{fig:3}  
\end{figure}

Although the ambiguity of ``Tr$^{-1}$'' precludes results from logically
forcing any single explanation, the existence of this ambiguity logically
does force something, namely a dynamic that continually extends statements
of results and explanations.  From the proofs in Ref.~\cite{aop05} and the
lattice structure of knob domains and detector domains follows an endless
open cycle of stating experimental results and explaining these results,
illustrated in Fig.~\ref{fig:3}.  This cycle operates in a context not
limited to theory but including the experimental endeavors that theory
describes.  Although in this writing we cannot reach beyond quantum
formalism to touch them, we have experiments in mind as a background
against which a statement of results implied by an explanation can be
judged and, if incompatible, rejected.

Picture a ``penguin'' toy walking down a slope with a rolling gait,
leaning left and swinging its right leg, then leaning right and swinging
its left leg, on and on in a cycle.  The swing of the left leg corresponds
to choosing a pair of inequivalent explanations (which are guaranteed
to exist as outlined in Sect.~\ref{sect:3}).  The swing of the right leg
corresponds to calculating the parametrized probability measures implied
by the natural extension of these explanations to a bigger knob domain,
resulting in conflicting predicted probabilities.  Experiments can
eliminate at least one such probability measure extended over the bigger
domain, but then the bigger knob domain serves as a base from which to
``swing the left leg'' and the cycle goes on, without end.  

The details of the cycle for metrically inequivalent parametrized density
operators are described elsewhere \cite{tyler}. To show how the expanding
cycle works when the metrically inequivalent elements of an explanation
are the POVMs, we need the following.  \vskip9pt

\noindent\textbf{Lemma}: For $M$ and $M'$ POVMs on a common knob domain
$\bm{K}$ (but possibly involving distinct detector domains and distinct
Hilbert spaces $\mathcal{H}$ and $\mathcal{H}'$, respectively,
MetDev$(M,M')\ne 0$ implies that
\begin{eqnarray}
\lefteqn{(\exists \bm{k}_1,\bm{k}_2\in \bm{K})\ \sup_{\sigma \in
  \mbox{DensOp}(\mathcal{H})}|\mbox{Tr}_\mathcal{H}\{\sigma
  [M(\bm{k},1)-M(\bm{k},1)]\}|}\qquad\qquad\quad\nonumber\\
&\mcl\ne \mcl&\sup_{\sigma' \in
  \mbox{DensOp}(\mathcal{H'})}|\mbox{Tr}_{\mathcal{H}'}\{\sigma'
  [M'(\bm{k},1)-M'(\bm{k},1)]\}|.
\end{eqnarray}
\vskip9pt
\noindent\textit{Proof}:
Given any bounded hermitian operator $A$ on a Hilbert space $\mathcal{H}$,
we have $\|A\|=\sup_{\|\ket{x}\|\le 1 }|\bra{x}A\ket{x}|$; that is, the
norm of a bounded hermitian operator equals its numerical
radius \cite{bhatia}.  It is easy to show that the numerical radius is
equal to $\sup_{\sigma\in
\mbox{DensOp}(\mathcal{H})}\mbox{Tr}_\mathcal{H}(\sigma A)$.  Thus we
have the following proposition:
\begin{equation}
  \|M(\bm{k}_1,\bm{\omega})- M(\bm{k}_2,\bm{\omega})\| = \sup_{\sigma\in
\mbox{DensOp}(\mathcal{H})}|\mbox{Tr}_\mathcal{H}\{\sigma [M(\bm{k}_1,\bm{\omega})-
M(\bm{k}_2,\bm{\omega})]\}|.
\end{equation}
From this the proof of the lemma follows. Q.E.D.  

Now we put the lemma to work.  A way to distinguish the POVM
$M(\bm{k}_1,-)$ from $M(\bm{k}_2,-)$ is to find the density operator
$\sigma$ that maximizes the separation of the probability measures ensuing
from these POVMs.  Because of the lemma, when MetDev$(M,M')\ne 0$, the
separation of probabilities ensuing from the parametrized POVM $M$ differs
from the separation of probabilities ensuing from $M'$.  Then if the
explanations involving $M$ and $M'$ are extended to a larger knob domain
$\bm{K}'$ that provides for $\sigma$ and $\sigma'$, respectively, there
results a conflict in the results implied by these extended explanations.
Once that happens, one rejects at least one of the explanations, say on
the basis of experiment, and keeps the other.  Then one treats the
extended knob domain $\bm{K}'$ as a new starting knob domain, and off we
go for another round of the cycle, as illustrated in Fig.~\ref{fig:3}.  A take-home
lesson is that a quantum description makes sense only as an element of a
family of related descriptions, and the related descriptions spread out of
larger and larger domains of knobs and detectors.

\subsection{The concept of invariance demands ambiguity of 
description}\label{subsect:4.3}

Now consider the concept of the invariances of a parametrized
probability measure under changes of knob settings.  

\noindent\textbf{Remark}: For those who like to think about Lorentz
invariance of electromagnetic theory, we note that in the quantum context,
the electromagnetic fields belong to the ``explanation'' part of the
story, and explanations need not be Lorentz invariant.  Indeed, neither
classically nor in quantum theory is the electromagnetic field a Lorentz
scalar.  It is not the field as an explanation but the probabilities of
detection, that are supposed to be the same for two experiments, one
conducted for example in the inertial frame of train station, the other in
the inertial frame of a train in uniform motion relative to the train
station.  Here we focus not on explanations but on ``statements
of results'' expressed as parametrized probability measures.

An invariance in a parametrized probability measure $\mu$
asserts a non-trivial equivalence class on its knob domain $\bm{K}$, an
equivalence class $[\bm{k}]$ defined by
\begin{equation}
[\bm{k}] = \{\bm{k}'|\mu(\bm{k}',-)=\mu(\bm{k},-)\}.
\end{equation}
An example is a violation of Bell inequalities by which entanglement is
demonstrated \cite{ppm07}.  In that example the probability of coincidence
detection by two rotatable detectors, one turned through an angle $k_A$,
the other through an angle $k_B$ is $\mu(k_A,k_B,(1,1))=
\half\cos^2(k_A-k_B)$. This and the other relevant probabilities
depend on $k_A$ and $k_B$ only as their difference $k_A-k_B$, so that
a change defined by adding the same amount of rotation to each of these
knobs leaves $\mu$ invariant.

But here is a conceptual muddle.  If changing the knob settings makes no
difference to the results, on what basis can we judge that any change in
knob settings has taken place?  A related question was put by one of our
mathematician colleagues: Why not just ``mod out'' the equivalence
classes?  But it won't do for physics to ``mod out'' such an equivalence
class; the physicist wants not to make it disappear but to appreciate it.

One way to appreciate changes of knobs that make no difference to the
results is to recognize, side by side with the statement of results $\mu$, a
second statement of results $\mu'$ at a finer level of detail, in
particular a
detector domain augmented by extra detectors to register changes in $k_A$
and $k_B$ separately.  Then $\mu$ is seen as obtained from $\mu'$
by ignoring the ``extra'' detectors:
\begin{equation}
  \mu'(k_A,k_B,(1,1,\Omega_{\rm knob}))= \mu(k_A,k_B,(1,1)).
\end{equation}
Here $\Omega_{\rm knob}$ is the ``anything-goes'' or ``don't care''
outcome of the extra detectors that respond to $k_A$ and $k_B$
separately, so that $\mu$ is seen as a \emph{marginal} probability measure
derived from ignoring ``knob-motion detectors'' in a more detailed
statement of results $\mu'$ that breaks invariance to show that $k_A$ and
$k_B$ moved even if their difference was held fixed.

A second way to make sense of invariance of results is to understand the
invariant parametrized probability measure $\mu$ over $\bm{K}$ as derived
from a second parametrized probability measure $\mu'$ over a larger knob
domain $\bm{K}'$ that contains an extra knob.  $\mu$ is then obtained from
$\mu'$ by fixing the extra knob at a special value.  

For example, to demonstrate rotational invariance we might place a disk on
a table and rotate it to show that ``nothing detectable changes under
rotation.''  But to see this invariance, whether one is aware of it or
not, one must manage incompatible frames of reference \cite{byers}.
Looked at one way ``nothing happens when we rotate the disk; but to see
that ``nothing happens when we rotate the disk'' one must see in the
other frame, so to speak, that in fact ``the disk rotates.'' 
This suggests adding a knob that can move the center of rotation away from
the center of the disk.  When the disk is off center, one sees its
rotation.  As the center of the disk is moved closer to the center of
rotation, one approaches invariance.

Something similar can be worked out for the preceding example involving
quantum states that violate Bell inequalities.  When this is done, the
equivalence class of knob settings show up as singular values \cite{sing}
in the mapping $\mu'$ from knobs to probability measures, leading to
another avenue for exploration.

\subsection{Remarks on quantum key distribution}\label{subsect:4.4}

Designs for quantum key distribution \cite{crypt} assert security against
undetected eavesdropping, based on transmitting quantum states that
overlap, with the result that deciding between them with neither error nor
an inconclusive result is impossible.  The most popular design,
BB84 \cite{BB84}, invokes four states (which we write as density operators)
$\rho(1),\ldots,\rho(4)$.  The claim of security invokes propositions such
as this: if $\frac{1}{2}\mbox{Tr}|\rho(1)-\rho(2)|\le \frac{1}{\sqrt{2}}$
then, by a well known result of quantum decision theory the least possible
probability of error to decide between them is:
\begin{eqnarray}
P_E&\mcl \ge \mcl&\textstyle{\frac{1}{2}}(1-
\textstyle{\frac{1}{2}}|\rho(1)-\rho(2)|)
=\textstyle{\frac{1}{2}}\left(1-\sqrt{\textstyle{\frac{1}{2}}}\right)
\approx 0.146 .
\label{eq:pe}
\end{eqnarray}

But how is one to rely on an implemented key-distribution system built
from lasers and optical fibers and so forth to act in accordance with this
explanation?  If a system of lasers and optical fibers and so forth
``possessed'' a single explanation in terms of quantum states, one could
hope to test experimentally the trace distance between the pair of states.
But no such luck.  The trouble is that trace distance is a property not of
probabilities \emph{per se}, which are testable, but of some one among the
many \emph{explanations} of those probabilities.  While the testable
probabilities constrain the possible explanations, and hence constrain
trace distances, this constraint on trace distance is ``the wrong way
around''---a lower bound instead of a sub-unity upper bound on which
security claims depend.    

Given any parametrized probability measure, proposition 2 in Ref.~\cite{aop05}
assures the existence of an explanation in terms of a parametrized density
operator $\rho'$ metrically inequivalent to $\rho$, such that, in
conflict with Eq.\ (\ref{eq:pe}), the trace distance becomes
$\textstyle{\frac{1}{2}}|\rho'(1)-\rho'(2)| = 1$, making the quantum states
in this explanation distinguishable without error, so that the keys that
they carry are totally insecure.  

The big question in key distribution is this: how will the lasers and
fibers and detectors that convey the key respond to attacks in which an
unknown eavesdropper brings extra devices with their own knobs and
detectors into contact with the key-distributing system?  Attacks entail
knob and/or detector domains extended beyond those tested, with the
possibility that extended explanations metrically inequivalent to that
used in the design, but consistent with available probabilities, both
imply a lack of security theoretically and accord with actual
eavesdropping.

Physically, one way for insecurity to arise is by an information leak
through frequency side-band undescribed in the explanation on which system
designers relied.  A more likely security hole appears when lasers that
are intended to radiate at the same light frequency actually radiate at
slightly different frequencies, as described in Refs.~\cite{ppm07,spie05,Frwk}.

\section{Concluding remarks}\label{sect:5}

\begin{figure}[t]
\centerline{\includegraphics[width=4.9275in]{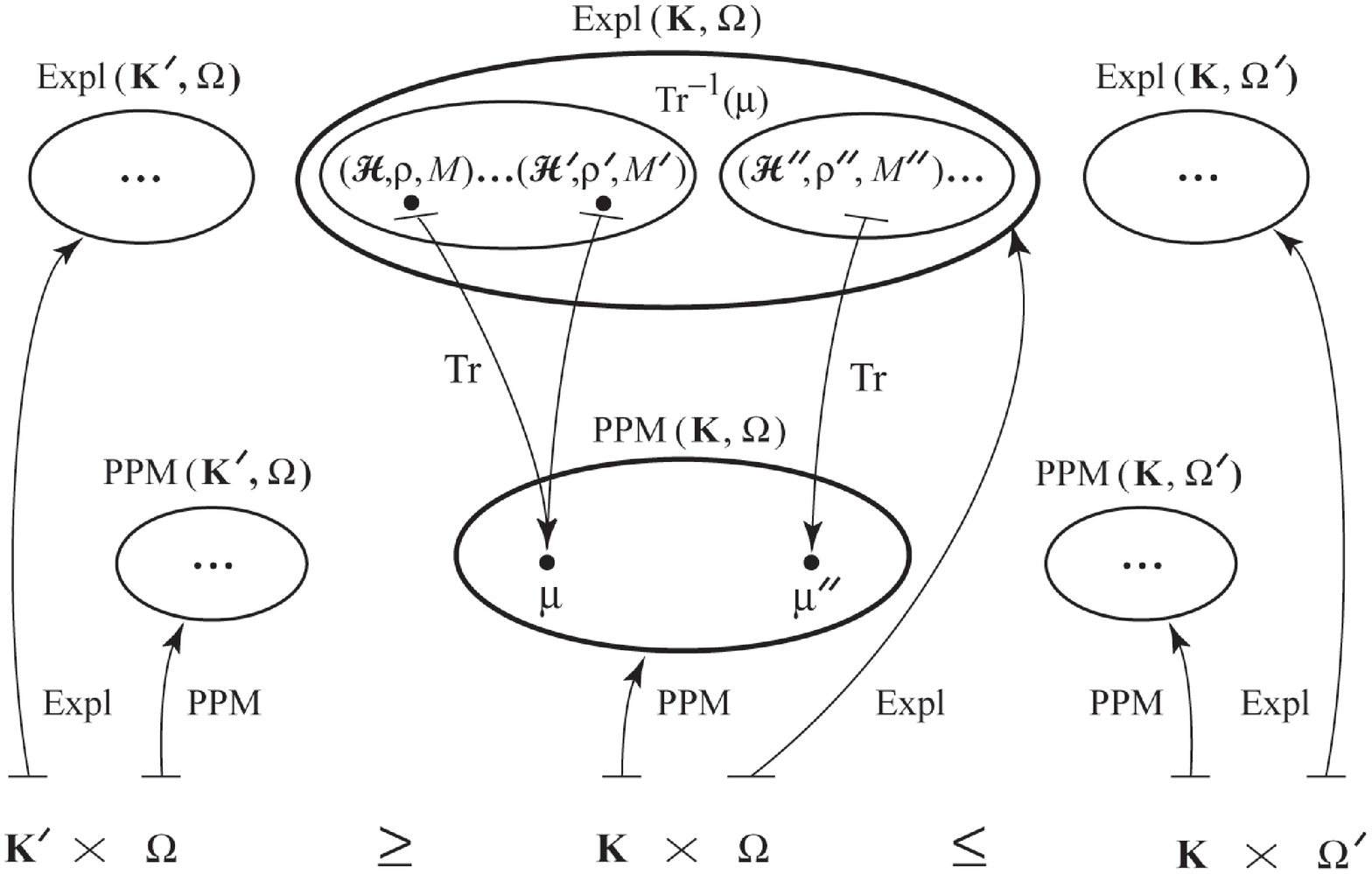}}
\caption{$\mathrm{Tr}^{-1}(\mu)$ contains many explanations.}\label{fig:4}
\end{figure}

As discussed in Sect.~\ref{subsect:4.2} and illustrated by Fig.~\ref{fig:3}, 
the roominess of the
inverse trace forces an open cycle of expanding descriptions, encompassing
both expansions of explanations and expansions of statements of results,
along with expansions of their knob domains and their outcome domains.
The discussion of invariance in Sect. 4.3 shows how understanding each
description as an element of a family of competing descriptions resolves
what otherwise is a conceptual obstacle. In the example of quantum key
distribution of Sect.~\ref{subsect:4.4}, we see how isolating a single description as if
competing descriptions were irrelevant confuses the role of quantum theory
in cryptography, with negative implications for the validity of claims of
security.  The world of multiple, competing descriptions in which quantum
engineering navigates is cartooned in Fig.~\ref{fig:4}.

\section*{Acknowledgments} We are grateful for helpful discussions
with Howard Brandt, Louis Kauffman, Samuel Lomonaco, and Ravi Rau.

\end{document}